\documentclass[12pt]{article}
\usepackage{amsfonts,amsmath,amssymb,epsfig,epsfig}
\renewcommand{\vec}[1]{\boldsymbol{#1}}

\begin{document}
%+++++++++++++++++++++++++++++++++++++++++++++++++++++++++++++++++++++
\title{\noindent \textbf{Explicit secular equations for piezoacoustic
surface waves: \\Rayleigh modes.\\
}}
%+++++++++++++++++++++++++++++++++++++++++++++++++++++++++++++++++++++

\author{Bernard Collet, Michel Destrade}
\date{2005}
\maketitle

\thispagestyle{empty}
%
%++++++++++++++++++++++++++++++++++++++++++++++++++++++++++
\begin{abstract}
{\footnotesize The existence of a two-partial Rayleigh wave
coupled to an electrical field in 2mm piezoelectric crystals is
known but has rarely been investigated analytically. It turns out
that the $Z$-cut $X$-propagation problem can be fully solved, up
to the derivation of the secular equation as a polynomial in the
squared wave speed. For the metallized (unmetallized) boundary
condition, the polynomial is of degree 10 (48). The relevant root
is readily identified and the full description of the mechanical
and electrical fields follows. The results are illustrated in the
case of the superstrong piezoelectric crystal, Potassium niobate,
for which the effective piezoelectric coupling coefficient is
calculated to be about 0.1}

\end{abstract}

%%++++++++++++++++++++++++++++++++++++++++++++++++++++++++++

\newpage

%++++++++++++++++++++++++++++++++++++++++++++++++++++++++++++++++++
\section{Introduction}
%++++++++++++++++++++++++++++++++++++++++++++++++++++++++++++++++++

This article prolongs and complements papers by the present
authors \cite{CoDe04} and by others \cite{Bleu68, Guly69, Tsen70,
KoVo73, BrGi79, BrHu89, MoWe02} where the propagation of a
Shear-Horizontal (SH) surface acoustic wave, decoupled from a
two-partial Rayleigh surface acoustic wave, was considered for
piezoelectric crystals. Those papers examined situations (cuts,
propagation directions) where the interaction between acoustic
fields and piezoelectric fields concerns the SH wave exclusively
and not the Rayleigh wave, which remains purely elastic. In the
present paper, the situation is reversed: the interaction occurs
solely between the electric field and the mechanical displacement 
lying in the sagittal plane (the plane containing
the direction of propagation and the direction of attenuation),
leading to a piezoacoustic two-partial (elliptically polarized) 
Rayleigh surface wave.

The properties of a two-partial Rayleigh surface wave  complement
those of a SH surface wave and one wave's loss is the other's
gain. Hence SH surface waves are particularly suited for immersed
crystals (liquid sensing, biosensors, etc.) because the mechanical
displacement is polarized horizontally with respect to the
interface, which leads to low loss of acoustic power in the fluid;
conversely, two-partial Rayleigh surface waves are used
extensively for non-destructive surface evaluation\cite{Vikt67}
and for free surface sensors\cite{Draf01}, because their
propagation is highly sensitive to anything present on the
interface which might perturb their vertical displacement. To take
but one example it is possible, using Rayleigh surface waves, to
design a mass microbalance with a mass resolution of 3
picograms\cite{BoCD91}.

This context reveals the importance of studying the analytical
properties of such waves. The cuts allowing for the propagation of
two-partial Rayleigh waves coupled to an electric field were
identified and classified by Maerfeld and Lardat \cite{MaLa70};
these waves were also investigated numerically \cite{CaJo70, MBDR71} 
and experimentally \cite{BRCD71}, as is best recalled in the textbook
by Royer and Dieulesaint \cite{RoDi00} (see also Mozhaev and
Weihnacht \cite{MoWe02} for pointers to more recent
contributions.) In general, the problem treatment however falls
short of a full analytical resolution, and the wave speed is
usually found from a trial-and-error procedure which goes back and
forth between the propagation condition and the boundary
condition, until a certain determinant is minimized to a required
degree of accuracy \cite{CoTi67} (alternatively, Abbudi and
Barnett \cite{AbBa90} proposed a numerical scheme based on the
surface-impedance matrix.) The present paper shows that a secular
equation can be derived explicitly as a polynomial of which the
wave speed is a root, for the $Z$-cut $X$-propagation problem.

This feat is achieved by use of some fundamental equations (II)
satisfied by the 6-vector whose components are the mechanical
displacements and tractions and the electrical potential and
induction at the interface. Albeit powerful, the method based on
the fundamental equations has one drawback because the polynomial
secular equation possesses several spurious roots. Hence for the
metallized boundary condition (III.A), it is a polynomial of
degree 10 in the squared wave speed, 
and for the unmetallized boundary condition (III.B), it
is a polynomial of degree 48! Nevertheless, finding the numerical
roots of a polynomial is almost an instantaneous process for a
computer. Also, it is expected that among all the 10 or 48
possible roots, one gives \textit{exactly} the surface wave speed.
Consequently, that root satisfies the boundary condition exactly,
whereas none of the spurious roots does. Once the relevant root is
thus properly identified, all the quantities of interest follow
naturally: the attenuation coefficients, the depth profiles, the
electromagnetic coupling coefficient, etc. Here, the method is
applied to the superstrong piezoelectric crystal, Potassium
niobate KNO$_3$, for which the effective electromagnetic coupling
coefficient for the piezoacoustic surface wave is found to be
about 0.1.

%++++++++++++++++++++++++++++++++++++++++++++++++++++++++++++++++++
\section{Basic equations}
%++++++++++++++++++++++++++++++++++++++++++++++++++++++++++++++++++

%------------------------------------------------
\subsection{Constitutive equations and equations of motion}
%------------------------------------------------

Consider a piezoelectric crystal with two mirror planes
(orthorhombic 2mm, tetragonal 4mm, or hexagonal 6mm).
For this type of crystal, the elasto-piezo-dielectric
matrix\cite{IEEE88} is of the form,
\begin{equation} \label{constitutive}
\begin{array}{|c c c c c c | c c c |}
   \hline
\bullet & \bullet & \bullet & & & & & & \bullet \\
\bullet & \bullet & \bullet & & & & & & \bullet \\
\bullet & \bullet & \bullet & & & & & & \bullet \\
        &         &         & \bullet & & & & \bullet & \\
        &         &         &  & \bullet & & \bullet  & & \\
        &         &         &  &         & \bullet & & & \\
  \hline
        &         &         &  &  \bullet &  & \bullet & & \\
        &         &         &  \bullet & &  & &  \bullet & \\
   \bullet & \bullet &  \bullet &  & &  & & &  \bullet\\
   \hline
\end{array}
\end{equation}

Now consider the $Z$-cut, $X$-propagation of a surface acoustic
wave that is, a motion with speed $v$ and wave number $k$ where
the displacement field $\vec{u}$ and the electric potential $\phi$
are of the form,
\begin{equation} \label{uPhi}
 \{ \vec{u}, \phi \} (x_1, x_2, x_3, t)
 = \{ \vec{U}(k x_3), \varphi(kx_3) \} \text{e}^{ik(x_1 - vt)},
\end{equation}
(say), with
\begin{equation} \label{BC1}
 \vec{U}(\infty) = 0, \quad \varphi(\infty) = 0.
\end{equation}
Here the $x_1$, $x_2$, $x_3$ axes are aligned with the 
crystallographic axes, and the crystal occupies the $x_3 \ge 0$ region.

It follows from the constitutive equation Eq.~\eqref{constitutive}
that the tractions $\sigma_{ij}$ and the electric induction $D_i$
are of a similar form,
\begin{equation} \label{wave2}
\{ \sigma_{ij}, D_i \} (x_1,x_2,x_3,t) =
 ik \{ t_{ij}(kx_3), d_i(kx_3) \} \text{e}^{ik (x_1 - vt)},
\end{equation}
(say) with $t_{22} = - ic_{23}U_3' - ie_{32} \varphi'
          + c_{12} U_1$, $d_2 = -ie_{24} U_2'$,
\begin{align}  \label{wave3}
&t_{11} = - ic_{13}U_3' - ie_{31} \varphi' + c_{11} U_1,
\nonumber \\
&t_{13} = - ic_{55} U_1' +  c_{55}U_3  +  e_{15}\varphi,
\nonumber \\
&t_{33} = - ic_{33}U_3' - ie_{33} \varphi' + c_{13} U_1,
\nonumber \\
&t_{23} = -ic_{44} U'_2 + c_{46}U_2, \quad
t_{12} = -ic_{46} U'_2 + c_{66}U_2,
\nonumber \\
&d_1= -ie_{15} U_1' + e_{11} U_1 + e_{15} U_3 - \epsilon_{11}\varphi,
\nonumber \\
&d_3= - ie_{33}U_3' + i\epsilon_{33} \varphi' + e_{31} U_1,
\end{align}
where the prime denotes differentiation with respect to $kx_3$.
Also, the surface wave vanishes away from the interface, so that
\begin{equation} \label{BC2}
 t_{ij}(\infty) = 0, \quad d_i(\infty) = 0.
\end{equation}

The classical equations of piezoacoustics, $\sigma_{ij,j} =  \rho
u_{i,tt}$, $D_{i,i} = 0$ (where $\rho$ is the mass density of the
crystal), reduce to
\begin{align}
& - t_{11} + i t'_{13} = -\rho v^2 U_1, \quad
 && - t_{12} + i t'_{23} = -\rho v^2 U_2,
\notag \\
& - t_{13} + i t'_{33} = -\rho v^2 U_3, \quad
 && - d_1 + i d'_3 = 0.
\label{piezoacoustics}
\end{align}

Clearly, the second equation Eq.~\eqref{piezoacoustics}$_2$ involves
only the function $U_2$ and is decoupled from the three others,
which involve the functions $U_1$, $U_3$, and $\varphi$.
It reads: $c_{44} U_2'' - (c_{66} - \rho v^2)U_2 = 0$.
A simple analysis shows that there are no functions $U_2$ solution
to this second-order ordinary differential equation such that
$U_2(\infty)=0$ and $t_{23}(0) =  -ic_{44} U'_2(0) = 0$,
except the trivial one.
Hence, the piezoelastic equations, coupled with free surface boundary
condition, lead to \textit{plane strain}: $U_2=0$, which in turn leads
to (generalized) \textit{plane stress}: $t_{12} = t_{23} = d_2 = 0$ by
Eqs.~\eqref{wave3}$_{4,6,8}$.

Now the remaining constitutive equations and piezoacoustic equations
can be arranged as a  first-order linear differential system.
It develops as:
\begin{equation}  \label{motion}
\mbox{\boldmath $\xi$}'
 = i \vec{N} \mbox{\boldmath $\xi$},
\end{equation}
where (using the notation of Ting\cite{Ting04})
\begin{equation}  \label{xi}
\mbox{\boldmath $\xi$}(kx_2) =
 \begin{bmatrix} U_1 \\ U_3 \\ \varphi \\ t_{31} \\ t_{33} \\ d_3
 \end{bmatrix},
\quad
 \vec{N} = \begin{bmatrix}
  0 & -1 & -s_6 & n_{66} & 0 & 0 \\
-r_4 & 0 & 0 & 0 & n_{22} & n_{24} \\
-r_2 & 0 & 0 & 0 & n_{24} & n_{44} \\
X-\eta & 0 & 0 & 0 & -r_4 & -r_2 \\
0 & X & 0 & -1 & 0 & 0 \\
0 & 0 & -\mu & -s_6 & 0 & 0
              \end{bmatrix}.
\end{equation}
Lothe and Barnett\cite{LoBa76} established the explicit
expressions for the components of the real matrix $\vec{N}$ in the
general case (general anisotropy, general piezoelectricity). In
the present context, they are given by
\begin{align}
& X = \rho v^2, \quad
      \delta^2 = \epsilon_{33} c_{33} + e_{33}^2,
\nonumber \\
& s_6 = e_{15}/c_{55},  \quad
  r_4 = (\epsilon_{33}c_{13} + e_{31}e_{33})/\delta^2, \quad
  r_2 = (c_{13}e_{33} - e_{31}c_{33})/ \delta^2,
\nonumber \\
& n_{66} = 1 / c_{55}, \quad
  n_{22} = \epsilon_{33}/ \delta^2, \quad
  n_{24} = e_{33}/ \delta^2, \quad
  n_{44} =  -c_{33}/ \delta^2,
\nonumber \\
& \eta = c_{11} -
 [c_{13}(\epsilon_{33}c_{13}+2e_{31}e_{33}) - c_{33}e_{31}^2]/\delta^2,
\quad
 \mu = -(\epsilon_{11} + e_{15}^2 / c_{55}).
\end{align}

%------------------------------------------------
\subsection{General solution}
%------------------------------------------------

The solution to the linear system with constant coefficients
Eq.~\eqref{motion} is of exponential form. Indeed, taking
$\mbox{\boldmath $\xi$}$ as: $\mbox{\boldmath $\zeta$}e^{ikqx_3}$
where $\mbox{\boldmath $\zeta$}$ is a constant vector and $q$ a
decay coefficient, leads to the eigenvalue problem: $(\vec{N} - q
\vec{I_6})\mbox{\boldmath $\zeta$}
 = \vec{0}$ where
$\vec{I_6}$ is the $6 \times 6$ identity matrix. Hence, $q$ is a
root (with positive imaginary part, to ensure decay) to the
\textit{propagation condition}: $\text{det}(\vec{N} - q \vec{I_6})
= 0$, which is a cubic for $q^2$,
\begin{equation} \label{bicubic}
  q^6 - \omega_4 q^4 + \omega_2 q^2 - \omega_0 = 0,
\end{equation}
where
\begin{align}
& \omega_4 = n_{22}X + n_{66}(X-\eta) - n_{44}\mu + 2r_2s_6 + 2r_4,
\notag \\
& \omega_2 = \{(X-c_{55})
                 [\epsilon_{33}(X-c_{11}) - e_{31}^2]
\notag \\
& \phantom{123456}
    - \epsilon_{11}[X(c_{33}+c_{55}) -c_{11}c_{33}
                                      + c_{13}(c_{13}+2c_{55})]
\notag \\
& \phantom{123456789012}
    - e_{15}[X(e_{15}+2e_{31}) + 2e_{33})
                   -2e_{33}c_{11} + 2(e_{15}+e_{31})c_{13})]\}
\notag \\
& \phantom{123456789012345678}
    /[c_{55}(\epsilon_{33} c_{33} + e_{33}^2)],
\notag \\
& \omega_0 = -(X-c_{11})
                 (X\epsilon_{11} - e_{15}^2 - \epsilon_{11}c_{55})
              /[c_{55}(\epsilon_{33} c_{33} + e_{33}^2)].
\end{align}

Here of course, it must be realized that the \textit{propagation
condition} Eq.~\eqref{bicubic} can be solved for $q$ only once the
speed of the surface wave (and hence $X=\rho v^2$) is known.
The next subsection and the next section show how $X$ can be found as
a root of the  \textit{secular equation}.
Once $X$ is known, the propagation condition gives six roots, out of
which only three are kept: $q_1$, $q_2$, $q_3$ say, the three roots
with positive imaginary roots ensuring exponential decay (if for a
given $X$, the propagation condition fails to deliver three such
roots, then no surface wave can propagate at speed $\sqrt{X / \rho}$.)

Let $\mbox{\boldmath $\zeta^1$}$,  $\mbox{\boldmath $\zeta^2$}$,
$\mbox{\boldmath $\zeta^3$}$ be the corresponding eigenvectors:
$\vec{N} \mbox{\boldmath $\zeta^i$}
  = q_i  \mbox{\boldmath $\zeta^i$}$
($i=1,2,3$), obtained for example as the third column of the
matrix adjoint to $\vec{N} - q_i \vec{I_6}$. Explicitly they are
of the form,
\begin{equation}
 \mbox{\boldmath $\zeta^i$}  =
 [ a_i,  b_i,
         \dfrac{e_{15}}{\epsilon_{33}} c_i, c_{55} f_i,
       c_{55} g_i , \epsilon_0 \dfrac{e_{15}}{\epsilon_{33}} h_i ]^T,
\end{equation}
where the non-dimensional quantities $a_i$, $g_i$, $h_i$
contain only even powers of $q$,
and the non-dimensional quantities $c_i$, $f_i$, $g_i$
contain only odd powers of $q$:
\begin{align} \label{components}
& \epsilon_{33}c_{33}a_i = -(\epsilon_{33}c_{33} + e_{33}^2)q^4
 + [\epsilon_{33}(X-c_{55}) - \epsilon_{11}c_{33}
        - 2e_{15} e_{33}]q^2
   \notag \\
& \phantom{123456789}
    +  \epsilon_{11}(X-c_{55}) - e_{15}^2,
\notag \\
& \epsilon_{33}c_{33}b_i =
     [(c_{13} + c_{55})\epsilon_{33} + e_{33}(e_{15} + e_{31})]q^3
      + [(c_{13}+c_{55})\epsilon_{11} + e_{15}(e_{15}+ e_{31})]q,
\notag \\
& e_{15}c_{33}c_i =
  - [c_{33}(e_{15} + e_{31}) + e_{33}(c_{13}+c_{55})]q^3
      + [e_{15}(X+c_{13}) + e_{31}(X-c_{55})]q,
\notag \\
& \epsilon_{33}c_{33}c_{55}f_i =
  - c_{55}(\epsilon_{33}c_{33} + e_{33}^2)q^5
\notag \\
& \phantom{12345}
    + [\epsilon_{33}c_{55}(X+c_{13}) - c_{33}(\epsilon_{11}c_{55}
   + e_{15}^2 + e_{15}e_{31}) + e_{33}(e_{15}c_{13}+e_{31}c_{55})]q^3
\notag \\
& \phantom{1234567}
         + [(\epsilon_{11}c_{55}+e_{15}^2)(X+c_{13})
                   + Xe_{15}e_{31}]q,
\notag \\
&   \epsilon_{33}c_{33}c_{55}g_i
= - c_{55}(\epsilon_{33}c_{33} + e_{33}^2)q^4
+ [(\epsilon_{33}c_{13} + e_{33}e_{31})(X - c_{55})
 \notag \\
& \phantom{123456789}
            + e_{15}e_{33}(X-c_{13}) + c_{33}(\epsilon_{11}c_{55}
   + e_{15}^2 + e_{15}e_{31})]q^2
\notag \\
& \phantom{1234567890123}
         + c_{13}[\epsilon_{11}(X-c_{55}) - e_{15}^2],
\notag \\
&  e_{15} \epsilon_0 c_{33}h_i =
  e_{15}(\epsilon_{33}c_{33} + e_{33}^2)q^4
   \notag \\
& \phantom{12345}
           - [\epsilon_{33}e_{15}(X + c_{13})
        - e_{33}(\epsilon_{11}c_{55}+e_{15}^2-e_{15}e_{31})
   + \epsilon_{11}(e_{31}c_{33}-e_{33}c_{13})]q^2
\notag \\
& \phantom{1234567}
         + e_{31}[\epsilon_{11}(X-c_{55}) - e_{15}^2].
\end{align}
Then the general solution to the equations of motion
Eqs.~\eqref{motion} is
\begin{equation} \label{generalSolution}
\mbox{\boldmath $\xi$}(kx_3)
 = \gamma_1 \mbox{\boldmath $\zeta^1$} e^{ikq_1 x_3}
 +  \gamma_2 \mbox{\boldmath $\zeta^2$} e^{ikq_2 x_3}
  +  \gamma_3 \mbox{\boldmath $\zeta^3$} e^{ikq_3 x_3},
\end{equation}
where $\gamma_1$, $\gamma_2$, $\gamma_3$,
are constants.

Depending on the type of boundary conditions, a given
homogeneous system of three linear equations for
$\gamma_1$, $\gamma_2$, $\gamma_3$ is derived.
The corresponding determinantal equation is the
\textit{boundary condition}.
In general for surface waves, the interface $x_3=0$ remains
free of tractions: $t_{31}(0) = t_{33}(0) = 0$.
From these two equations, $\gamma_2$ and $\gamma_3$
can be expressed in terms of $\gamma_1$ as
\begin{equation} \label{normalize}
\dfrac{\gamma_2}{\gamma_1} =
 \dfrac{f_3g_1 - f_1g_3}{f_2g_3 - f_3g_2},
\quad
\dfrac{\gamma_3}{\gamma_1} =
 \dfrac{f_1g_2 - f_2g_1}{f_2g_3 - f_3g_2}.
\end{equation}

To sum up: first the speed of the surface wave must be
computed as a root of the \textit{secular equation} (Section III),
obtained thanks to the \textit{fundamental equations} presented below
(Section II.C).
Next the appropriate decay coefficients are computed as roots with
positive imaginary parts from the \textit{propagation condition}
Eq.~\eqref{bicubic}.
Then it must be checked that the \textit{boundary condition}
(Section III) is indeed satisfied.
If it is, then the \textit{complete solution} is given by
Eqs.~\eqref{uPhi}, \eqref{xi}$_1$, \eqref{generalSolution},
\eqref{normalize}.

%------------------------------------------------
\subsection{Fundamental equations}
%------------------------------------------------

Now some fundamental equations are presented, from which the
secular equation is found. Their derivation is short and is given
in Refs.~\cite{Dest03d, Dest04a, Dest04b}; they represent a
generalization to interface waves of works by Currie\cite{Curr79}
and by Taziev\cite{Tazi89} for elastic surface waves (see also
Ting\cite{Ting04} for a review.) They read
\begin{equation} \label{fundamental}
\overline{\mbox{\boldmath $\xi$}}(0)
 \cdot \vec{M^{(n)}}
   \mbox{\boldmath $\xi$}(0) = 0,
\quad \text{where} \quad
 \vec{M^{(n)}} =
   \begin{bmatrix} \vec{0} & \vec{I_3} \\
                            \vec{I_3} & \vec{0}
   \end{bmatrix} \vec{N}^n,
\end{equation}
and $n$ is any positive or negative integer. By computing the
integer powers $\vec{N}^n$ of $\vec{N}$ (at $n=-2,-1,1,2,3$, say),
it is a simple matter to check that the $6 \times 6$ matrix
$\vec{M^{(n)}}$ is symmetric and that its form depends on the
parity of $n$. Hence $\vec{M^{(n)}}$ is of the forms,
\begin{equation} \label{2forms}
 \begin{bmatrix}
   0 & * & * & *  & 0 & 0\\
   * & 0 & 0 & 0 & * & * \\
  * & 0 & 0 & 0 & * & * \\
  * & 0 & 0 & 0 & * & * \\
 0 & * & * & * & 0 & 0 \\
 0 & * & * & * & 0 & 0
 \end{bmatrix},
\quad
 \begin{bmatrix}
    * & 0 & 0 & 0 & * & * \\
 0 & * & * & * & 0 & 0\\
 0 & * & * & * & 0 & 0\\
 0 & * & * & * & 0 & 0\\
  * & 0 & 0 & 0 & * & * \\
  * & 0 & 0 & 0 & * & *
 \end{bmatrix},
 \end{equation}
when $n=-2,2$, and when $n=-1,1,3$, respectively.

%++++++++++++++++++++++++++++++
\section{$Z$-cut, $X$-propagation}
%++++++++++++++++++++++++++++++

%------------------------------------------------
\subsection{Metallized boundary condition}
%------------------------------------------------

For metallized  (short-circuit) boundary conditions, the mechanically
free interface $x_3=0$ is covered with a thin metallic film,
grounded to potential zero, and so
\begin{equation} \label{BCmetal}
\mbox{\boldmath $\xi$}(0)
 = \gamma_1
  \begin{bmatrix}
     a_1 \\
     b_1 \\
     \dfrac{e_{15}}{\epsilon_{33}}c_1 \\
     c_{55} f_1 \\
     c_{55} g_1 \\
     \epsilon_0 \dfrac{e_{15}}{\epsilon_{33}}h_1
  \end{bmatrix} +
 \gamma_2
  \begin{bmatrix}
    a_2 \\
    b_2 \\
    \dfrac{e_{15}}{\epsilon_{33}}c_2 \\
    c_{55} f_2 \\
    c_{55} g_2 \\
    \epsilon_0  \dfrac{e_{15}}{\epsilon_{33}}h_2
  \end{bmatrix}+
\gamma_3
  \begin{bmatrix}
    a_3 \\
    b_3 \\
    \dfrac{e_{15}}{\epsilon_{33}}c_3 \\
    c_{55} f_3 \\
    c_{55} g_3 \\
    \epsilon_0 \dfrac{e_{15}}{\epsilon_{33}}h_3
  \end{bmatrix}
  =
  \begin{bmatrix}
    U_1(0) \\
    U_3(0) \\
    0 \\
    0 \\
    0 \\
    d_3(0)
  \end{bmatrix}.
\end{equation}

Two possibilities arise for the roots
with positive imaginary part of the bicubic Eq.~\eqref{bicubic}.
Either \textbf{(a)} $q_i = i \hat{q}_i$ ($\hat{q}_i > 0$)
or \textbf{(b)} $q_1 = -\overline{q_2}$,
$q_3 = i \hat{q}_3$ ($\hat{q}_3 > 0$).
In case \textbf{(a)}, it is clear from Eqs.~\eqref{components}
that $a_i$, $g_i$, $h_i$, are real numbers and that
$b_i$, $c_i$, $f_i$ are pure imaginary numbers.
Then, separating the real part from the imaginary part in the
third, fourth, and fifth lines in Eq.~\eqref{BCmetal}$_2$, it is found
that $[\gamma_1, \gamma_2, \gamma_3]^T$ is parallel to a real
vector.
It follows from Eq.~\eqref{BCmetal}$_1$ that
$\mbox{\boldmath $\xi$}(0)$ is of the form
\begin{equation} \label{xiMetal}
\mbox{\boldmath $\xi$}(0) =
 U_1(0)[1, i\alpha_2, 0, 0, 0, \beta_1]^T,
\end{equation}
where $i\alpha_2 := U_3(0)/U_1(0)$ is pure imaginary
($\alpha_2$ is real) and
$\beta_1 := d_3(0)/U_1(0)$ is real.
In case \textbf{(b)} a slightly lengthier study shows that
$ \mbox{\boldmath $\xi$}(0)$ is also of this form (see
Ting\cite{Ting04} and Destrade\cite{Dest04b} for proofs in different,
but easily transposed, contexts.)

Now substituting this expression Eq.~\eqref{xiMetal} for
$\mbox{\boldmath $\xi$}(0)$ into the fundamental equations
Eq.~\eqref{fundamental}$_1$ leads to a trivial identity when $n=-2,2$,
and to the following set of three equations when $n=-1,1,3$,
\begin{equation} \label{systemMetal}
\begin{bmatrix}
 M^{(-1)}_{22} &  M^{(-1)}_{16} & M^{(-1)}_{66} \\
 M^{(1)}_{22} &  M^{(1)}_{16} & M^{(1)}_{66} \\
 M^{(3)}_{22} &  M^{(3)}_{16} & M^{(3)}_{66}
\end{bmatrix}
  \begin{bmatrix} \alpha_2^2 \\ 2\beta_1 \\ \beta_1^2
  \end{bmatrix}
= \begin{bmatrix}
      - M^{(-1)}_{11} \\ - M^{(1)}_{11} \\ -M^{(3)}_{11}
   \end{bmatrix}.
\end{equation}
Note that the components of the $3 \times 3$ matrix and of
the right hand-side column vector above are easily
computed from their definition Eq.~\eqref{fundamental}$_2$;
for instance, $M^{(1)}_{22} =X$, $M^{(1)}_{16} = -r_6$,
$M^{(1)}_{66}=n_{44}$, $M^{(1)}_{11}=X-\eta$.

Cramer's rule applied to the system above reveals that
$2\beta_1 = \Delta_2 / \Delta$, $\beta_1^2 = \Delta_3 / \Delta$
(where $\Delta$ is the determinant of the $3 \times 3$ matrix in
Eq.~\eqref{systemMetal} and $\Delta_2$, $\Delta_3$ are the determinants
of the matrix obtained from this matrix by replacing the 2$^\text{nd}$
and 3$^\text{rd}$ columns by the right hand-side column in
Eq.~\eqref{systemMetal}, respectively) and so, that
\begin{equation} \label{seculMetal}
\Delta_2^2 - 4 \Delta \Delta_3 = 0.
\end{equation}
This is the \textit{explicit secular equation for the speed of a
two-partial Rayleigh piezoacoustic surface wave propagating in
a metallized mm2 (or 4mm, or 6mm) crystal}.

Its expression is too lengthy to reproduce here but has been
obtained using Maple.
It turns out that the secular equation is a polynomial of degree
10 in $X$.
Note also that the solution to the system Eq.~\eqref{systemMetal}
for the unknown $\alpha_2^2$ plays no role in the final expression of
the secular equation.
Hence the equation is only valid in the presence of piezoelectric
coupling through the solutions of Eq.~\eqref{systemMetal} for
$2\beta_1 (= 2d_3(0)/U_1(0))$ and for $\beta_1^2$, and it does not
cover the Rayleigh cubic function for purely elastic surface waves
in orthorhombic crystals.
Moreover, in the present context the in-plane piezoacoustic surface
wave is entirely decoupled from its anti-plane counterpart
(which does not exist, as seen in II.A) ,
and so the secular equation Eq.~\eqref{seculMetal}
cannot cover the case of a Bleustein-Gulyaev SH wave.
Hence in many respects, this new secular equation
is unique and stands alone, with no link whatsoever
with previously established secular equations.

Selecting the correct root or roots out of the 10 possible given by
the secular equation is quite a simple matter.
First the root $X$ must be real and positive;
then it must be such that the propagation condition Eq.~\eqref{bicubic}
written at $X$ yields three roots $q_1$, $q_2$, $q_3$ with positive
imaginary part; finally it must be such that the boundary conditions
Eqs.~\eqref{BCmetal}$_2$ are satisfied, that is
\begin{equation} \label{checkBCmetal}
\dfrac{1}{(q_1-q_2)(q_2-q_3)(q_3-q_4)}
\begin{vmatrix}
 c_1 & c_2 & c_3 \\
 f_1 & f_2 & f_3 \\
 g_1 & g_2 & g_3
\end{vmatrix} =0.
\end{equation}

For Potassium niobate\cite{ZSBV93} (KNbO$_3$, 2mm), 
the relevant constants are the following. 
Elastic constants ($10^{11}$ N m$^{-2}$): $c_{11} =
2.26$, $c_{13}=0.68$, $c_{33}=1.86$, $c_{55} = 0.25$;
Piezoelectric constants (C m$^{-2}$): $e_{15}= 5.16$,
$e_{31}=2.46$, $e_{33}=4.4$; Dielectric constants ($10^{-12}$ F
m$^{-1}$): $\epsilon_{11} = 34 \epsilon_0$, $\epsilon_{33} = 24
\epsilon_0$, $\epsilon_0 = 8.85416$ ; 
Mass density (kg m$^{-3}$): $\rho = 4630$. 
The secular equation has six complex roots and four real positive roots
in $X$; out of these four, only two are
such that the propagation condition yields suitable attenuation
coefficients; out of these two, only one is such that the boundary
condition is verified. The numerical values for the wave speed 
$V_\text{m}$ (say) and
for the attenuation coefficients are listed on the second line of
Table I. A 8 digit precision is given, although the calculations
were conducted with a 30 digit precision; the left hand-side in
Eq.~\eqref{checkBCmetal} was found to be smaller than $5 \times
10^{-22}$. 
The complete solution is found by taking the real part of 
the right hand-side in Eq.~\eqref{uPhi} and Eq.~\eqref{wave2}.
Specifically, the mechanical displacement $u_1$ is in
phase quadrature with the mechanical displacement $u_3$ 
and the electric potential,
\begin{align}
 & u_1 = \hat{U}_1(kx_3) \cos k(x_1 - vt), 
\notag \\
 & u_3 = \hat{U}_3(kx_3) \sin k(x_1 - vt), 
\notag \\ 
 & \phi = \hat{\varphi}(kx_3) \sin k(x_1 - vt), 
\end{align}
where $\hat{U}_1 := \Re\{U_1\}$, $\hat{U}_3 := \Im\{U_3\}$,
 $\hat{\varphi} := \Im\{\phi\}$ are the \textit{amplitude
functions}.
Figure 1 shows their variations with the scaled depth
$x_3/\lambda$, where $\lambda = 2\pi/k$ is the wavelength. The
vertical scaling is such that $U_1(0) = 1$\AA. 
The axes of the polarization ellipse are along the $x_1$ and $x_3$ 
axes. At the interface, $\hat{U}_1(0) > 0$,  $\hat{U}_3(0) < 0$, 
and  $|\hat{U}_3(0)| > |\hat{U}_1(0)|$, so that the major axis 
of the ellipse is along $x_3$ and the minor axis is along $x_1$;
there, the ellipse is spanned in the retrograde sense with time. 
The ellipse becomes more and more oblong with depth, and is linearly
polarized at a depth of about 0.174$\lambda$. 
Further down the substrate, it becomes elliptically polarized again, 
but is now spanned in the direct sense. 
It is circularly polarized at a depth of about 1.183$\lambda$, 
and then again linearly polarized at a
depth of about 0.987$\lambda$.

%------------------------------------------------
\subsection{Unmetallized boundary condition}
%------------------------------------------------

For the unmetallized (free) boundary condition, the free surface
is in contact with the vacuum (permeability: $\epsilon_0$), and
so\cite{CoDe04},
\begin{equation} \label{BCfree}
\mbox{\boldmath $\xi$}(0)
 = \gamma_1
  \begin{bmatrix}
    a_1 \\
    b_1 \\
    \dfrac{e_{15}}{\epsilon_{33}} c_1 \\
    c_{55} f_1 \\
    c_{55} g_1 \\
    \epsilon_0 \dfrac{e_{15}}{\epsilon_{33}} h_1
  \end{bmatrix} +
 \gamma_2
  \begin{bmatrix}
    a_2 \\
    b_2 \\
    \dfrac{e_{15}}{\epsilon_{33}} c_2 \\
    c_{55} f_2 \\
    c_{55} g_2 \\
    \epsilon_0\dfrac{e_{15}}{\epsilon_{33}} h_2
  \end{bmatrix}+
\gamma_3
  \begin{bmatrix}
    a_3 \\
    b_3 \\
    \dfrac{e_{15}}{\epsilon_{33}}c_3 \\
    c_{55} f_3 \\
    c_{55} g_3 \\
    \epsilon_0\dfrac{e_{15}}{\epsilon_{33}} h_3
  \end{bmatrix}
  =
  \begin{bmatrix}
    U_1(0) \\
    U_3(0) \\
    \varphi(0) \\
    0 \\
    0 \\
    i\epsilon_0 \varphi(0))
  \end{bmatrix}.
\end{equation}

Similarly to the previous case, the form of $\mbox{\boldmath $\xi$}(0)$
can be found, whatever the form of the $q_i$ is.
Namely,
 \begin{equation} \label{xiFree}
\mbox{\boldmath $\xi$}(0) =
 \varphi(0)[i\alpha_2, \beta_1, 1, 0, 0, i\epsilon_0]^T,
\end{equation}
where $i\alpha_2 = U_1(0)/\varphi(0)$ is pure imaginary
($\alpha_2$ is real) and
$\beta_1 = U_3(0)/\varphi(0)$  is real.
Substitution into the fundamental equations Eqs.~\eqref{fundamental}
at $n = -2,2$ leads to the trivial identity.
At $n=-1,1,3$ it leads to
\begin{equation} \label{systemFree}
M^{(n)}_{33} + \epsilon_0^2 M^{(n)}_{66}
 + 2\epsilon_0 M^{(n)}_{16}\alpha_2
  + 2 M^{(n)}_{23} \beta_1
   + M^{(n)}_{11} \alpha_2^2
    + M^{(n)}_{22} \beta_1^2 = 0,
\end{equation}
which are three equations for two unknowns $\alpha_2$ and
$\beta_1$.
Formally, solving two equations and substituting the result into the
third equation yields the secular equation.
Note however that these equations Eqs.~\eqref{systemFree} are
nonlinear (quadratic) in the unknowns.
Their resolution is somewhat lengthy, although possible as is now seen.

First take advantage of the identity $M^{(1)}_{23} \equiv N_{56} = 0$
to solve Eqs.~\eqref{systemFree}  at $n=1$ for $\beta_1^2$:
\begin{align}
\beta_1^2 & =
-[M^{(1)}_{33} + \epsilon_0^2 M^{(1)}_{66}
   + 2\epsilon_0 M^{(1)}_{16}\alpha_2
    + M^{(1)}_{11}  \alpha_2^2 ]/M^{(1)}_{22}
\notag \\
& = [\mu - n_{44} \epsilon_0^2  + 2 r_2 \epsilon_0 \alpha_2
         + (\eta - X) \alpha_2^2)]/X.
\end{align}
Next, solve Eqs.~\eqref{systemFree} at $n=-1,3$ for $\beta_1$:
\begin{equation}
-2 \beta_1 =
[M^{(n)}_{33} + \epsilon_0^2 M^{(n)}_{66}
  + 2\epsilon_0 M^{(n)}_{16}\alpha_2
   + M^{(n)}_{11}\alpha_2^2
    + M^{(n)}_{22} \beta_1^2]/M^{(n)}_{23},
\end{equation}
$n=-1,3$.
Now square both sides and substitute the expression for
$\beta_1^2$ just obtained to derive two polynomials of fourth
degree in $\alpha_2$.
Having $\alpha_2$ as a common root, these two polynomials
have a resultant equal to zero, a condition which is the
\textit{explicit secular equation for the speed of a
two-partial Rayleigh piezoacoustic surface wave propagating in
an unmetallized 2mm (or 4mm, or 6mm) crystal}.

Of course, the resulting polynomial is rather formidable, here of
degree 48 in $X$ according to Maple. Nevertheless, finding
numerically the roots of a polynomial is a quasi-instantaneous
task for a computer. For instance in the case of KNbO$_3$, it is
found that there are 10 positive real roots in $X$ to the
polynomial, out of which 6 yield three attenuation factors with
positive imaginary part. Out of these 6, only one satisfies the
boundary conditions Eqs.~\eqref{BCfree}$_2$, that is
\begin{equation} \label{checkBCfree}
\dfrac{1}{(q_1-q_2)(q_2-q_3)(q_3-q_4)}
\begin{vmatrix}
  f_1 & f_2 & f_3 \\
 g_1 & g_2 & g_3 \\
 h_1 - ic_1 & h_2 - ic_2 & h_3 - ic_3
\end{vmatrix} =0.
\end{equation}
Hence, at the speed $V_\text{u}$ (say)
and attenuation factors listed on the third
line of Table I (obtained with a 40 digits precision), the
determinant in Eq.~\eqref{checkBCfree} was found to be smaller
than $5 \times 10^{-23}$. Note by comparison of the second line and 
the third line of Table I that when the free surface is
metallized, the wave propagates at a slower speed, and is slightly
more localized, than when the surface is unmetallized. Figure 2
shows the variations of the amplitude functions 
with the scaled depth $x_3/\lambda$ in the unmetallized boundary 
conditons case. 
The depth curves are similar to those in the metallized case, 
with the differences that the boundary condition forces the 
electrical potential to be about 0.596 V at the interface, 
and that the nature of the polarization ellipse changes at depths 
which are slightly less than the corresponding depths with metallized 
boundary conditions.

Finally, recall that it is usual to take the quantity 
$2 (V_\text{u} - V_\text{m})/V_\text{u}$ 
as a measure of the crystal's ability to transform an electric signal 
into an elastic surface wave by means of interdigital electrode 
transducers although, as proved by Royer and Dieulesaint 
\cite{RoDi00}, the demonstration is far from obvious. 
This quantity is often referred to as the 
\textit{effective piezoacoustic coupling coefficient for surface 
waves} and is expected to be positive. 
In the present example of KNbO$_3$, the speeds of the second 
column in Table I give a value of 0.1037, far greater than the 
corresponding values\cite{CaJo70} 
for GaAS, Bi$_{12}$GeO$_{20}$, ZnO, and CdS, and more than twice that 
\cite{Farn78} for LiNbO$_3$.
Note that Mozhaev and Weihnacht \cite{MoWe00} 
reported negative values for this 
quantity corresponding to special cuts and propagation direction in 
KNbO$_3$. 

%++++++++++++++++++++++++++++++++++++++++++++++++++++++
% bibliography
%++++++++++++++++++++++++++++++++++++++++++++++++++++++

%%%%%%%%%%%%%%

\bigskip

\begin{center}
Table I. Wave speed (m s$^{-1}$) and attenuation coefficients for
a two-partial piezoacoustic surface wave in KNbO$_3$.

\noindent {\small
\begin{tabular}{l c c c}
\hline \rule[-3mm]{0mm}{8mm}
     & $V$ & $q_{1,2 }$  & $q_3$
\\
\hline metallized     & 3762.50953 &  $\pm$ 0.39191249 + $i$
0.49991830 & $i$ 3.10691826
\\
unmetallized &    3968.28624 & $\pm$ 0.39840475  + $i$ 0.45628905
& $i$ 3.07008806
\\
 \hline
\end{tabular}
}
\end{center}

\newpage 

\begin{figure}
 \centering 
  \mbox{\epsfig{figure=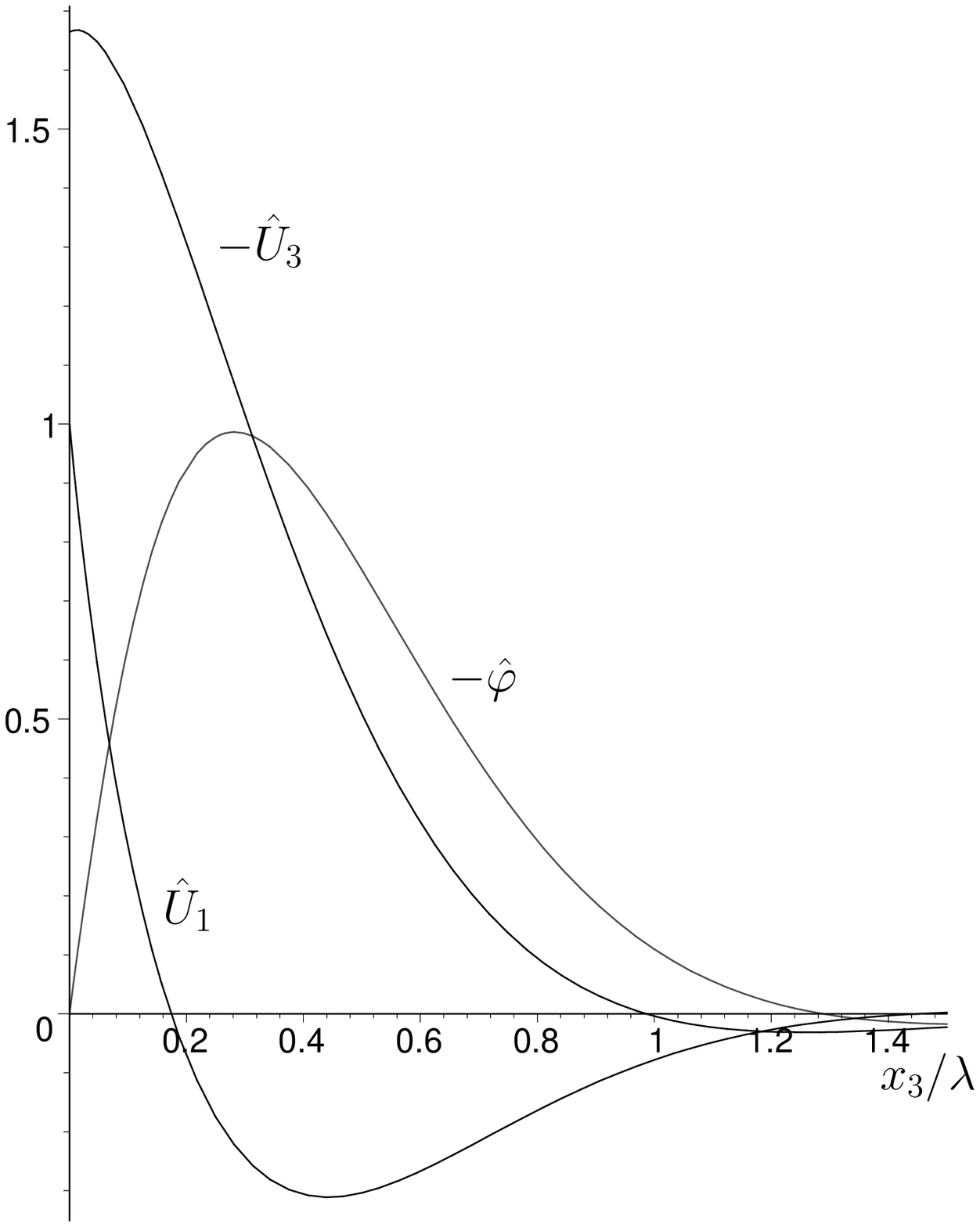}}
\caption{Depth profiles of the mechanical displacements (\AA)
  and the electric potential (V) for the piezoelectric Rayleigh wave 
   in KNbO$_3$,  $Z$-cut $X$-propagation 
   with metallized boundary conditions.}
\end{figure}

\begin{figure}
 \centering 
  \mbox{\epsfig{figure=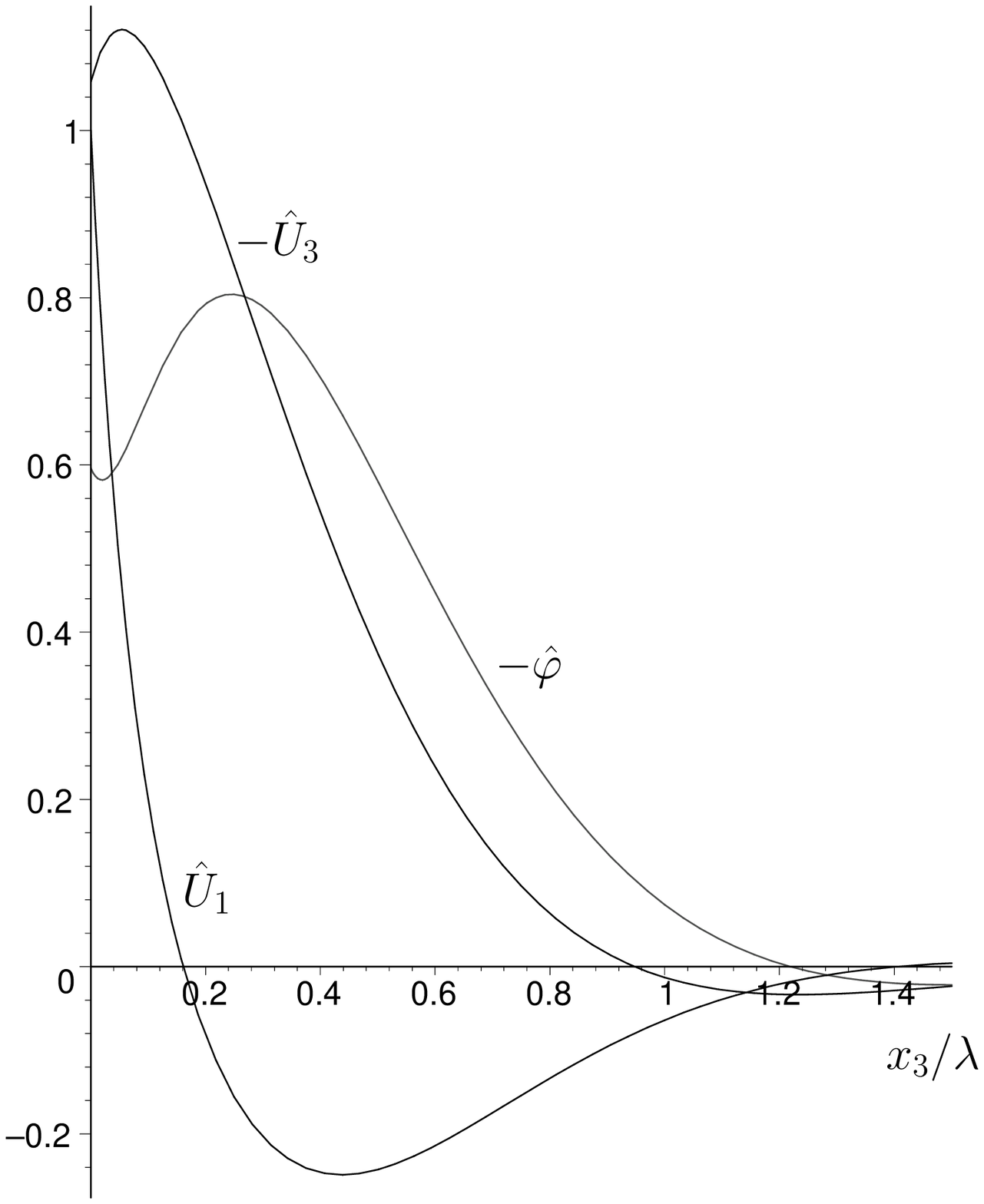}}
\caption{Depth profiles of the mechanical displacements (\AA)
  and the electric potential (V) for the piezoelectric Rayleigh wave 
   in KNbO$_3$, $Z$-cut $X$-propagation 
   with unmetallized boundary conditions.}
\end{figure}

%---------------------------------------------------------------

\end{document}